\documentclass[11pt,a4paper]{article}
\pdfoutput=1
\usepackage{jcappub}
\usepackage{bm}
\usepackage{amsfonts}
\usepackage{subfigure}

\newcommand{\M}[1]{#1}

\renewcommand\({\left(}
\renewcommand\){\right)}

\newcommand{\dPo}{\dot{P}_{\rm obs}}
\newcommand{\dPt}{\dot{P}_{\rm th}}

\newcommand{\Msun}{M_{\odot}}
\newcommand{\Lsun}{L_{\odot}}

\newcommand{\Nexp}{N_{\rm exp}}
\newcommand{\Napp}{\tilde{N}_{\rm exp}}

\begin{document}



\hfill DESY 15-245

\title{Cool WISPs for stellar cooling excesses}

\author[a]{Maurizio Giannotti}
\author[b]{Igor Irastorza}
\author[b,c]{Javier Redondo}
\author[d]{Andreas~Ringwald}

\affiliation[a]{Physical Sciences, Barry University,
11300 NE 2nd Ave., Miami Shores, FL 33161, USA}
\affiliation[b]{Departamento de F\'isica Te\'orica, Universidad de Zaragoza, Pedro Cerbuna 12, E-50009, Zaragoza, Espa\~na}
\affiliation[c]{Max-Planck-Institut f\"ur Physik (Werner-Heisenberg-Institut), F\"ohringer Ring 6, D-80805 M\"unchen, Germany}
\affiliation[d]{Theory group, Deutsches Elektronen-Synchrotron DESY\\Notkestra\ss{}e 85, D-22607 Hamburg, Germany}

\emailAdd{mgiannotti@barry.edu}
\emailAdd{igor.irastorza@cern.ch}
\emailAdd{jredondo@unizar.es}
\emailAdd{andreas.ringwald@desy.de}

\abstract{
Several stellar systems (white dwarfs, red giants, horizontal branch stars and possibly the neutron star in the supernova remnant Cassiopeia A) show a mild preference for a non-standard cooling mechanism when compared with theoretical models. This exotic cooling could be provided by Weakly Interacting Slim Particles (WISPs), produced in the hot cores and abandoning the star unimpeded, contributing directly to the energy loss. Taken individually, these excesses do not show a strong statistical weight. However, if one mechanism could consistently explain several of them, the hint could be significant. 
We analyze the hints in terms of neutrino anomalous magnetic moments, minicharged particles, hidden photons and axion-like particles (ALPs).  Among them, the ALP \M{or a massless HP } represent the best solution. Interestingly, the hinted ALP parameter space is accessible to the next generation proposed ALP searches, such as ALPS II and IAXO \M{and the massless HP requires a multi TeV energy scale of new physics that might be accessible at the LHC.}
}
\maketitle

\section{Introduction}
\begin{figure}[t]
	\centering
	\includegraphics[width=0.5\linewidth]{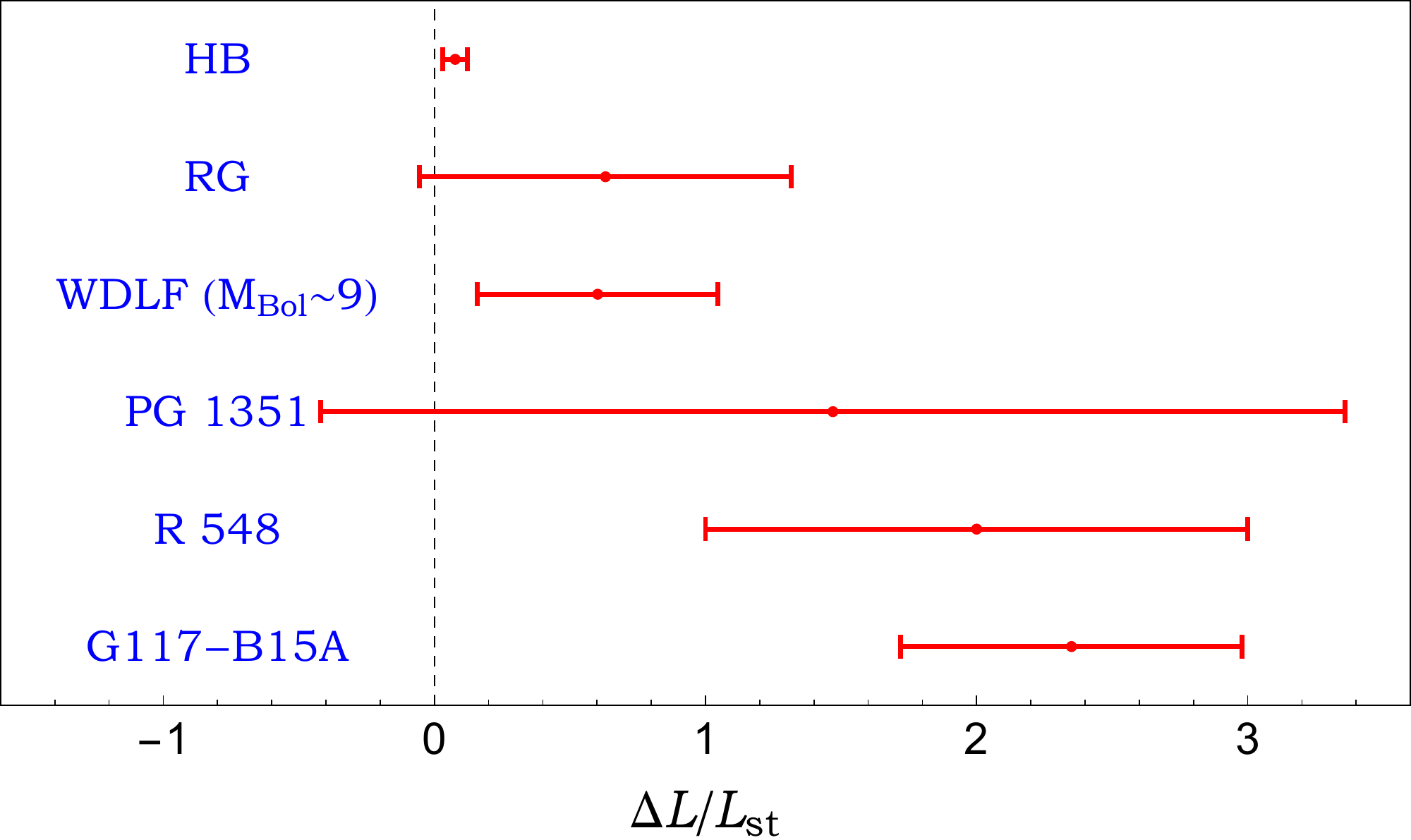}
	\caption{
		Missing energy loss, $ \Delta L $, normalized over a reference luminosity, $ L_{\rm st} $, for different stellar systems. 
		The plot includes only stars for which an analysis with confidence levels was provided: 
		the three white dwarf variables G117-B15A~\cite{Corsico:2012ki}, R548~\cite{Corsico:2012sh} and PG 1351+489~\cite{Corsico:2014mpa}; 
		an example from the central region of the WDLF ($ M_{\rm Bol} \sim 9$)~\cite{Bertolami:2014wua,Bertolami:2014noa}; 
		red giants~\cite{Viaux:2013hca,Viaux:2013lha}; and HB stars~\cite{Ayala:2014pea}.
		For RG and HB stars, the reference luminosity is taken to be the core average energy loss.
		The errors are derived from the $ 1\,\sigma $ uncertainties provided in the original literature. }
	\label{fig:WisePlot}
\end{figure}

For over two decades, observations of different stellar systems have shown deviations from the expected behavior, indicating in all cases an over-efficient cooling~\cite{Ringwald:2015lqa,Giannotti:2015dwa}. 
Deviations from the standard model expectations were found in very diverse stellar systems: 
white dwarfs (WDs)~\cite{KeplerEtAl,Isern:1992gia,BischoffKim:2007ve,Corsico:2012ki,Corsico:2012sh,Corsico:2014mpa,Bertolami:2014wua,Bertolami:2014noa}, old stars with inactive degenerate core; 
red giants (RGs)~\cite{Viaux:2013hca,Viaux:2013lha}, with a dense, degenerate core and a nondegenerate 
burning shell;
horizontal branch (HB) stars~\cite{Ayala:2014pea}, with a nondegenerate helium burning core;
helium burning supergiants~\cite{Skillman:2002aa,McQuinn:2011bb}, a few times more massive than our sun and with a hot and relatively low dense core;
and finally, very dense and hot neutron stars (NS)~\cite{Shternin:2010qi,Leinson:2014cja,Leinson:2014ioa}. 
Although each deviation, taken independently, has a small statistical significance, normally about $ 1\,\sigma $, together they overwhelmingly seem to indicate a systematic tendency of stars to cool more efficiently than predicted, as evident from Fig.~\ref{fig:WisePlot}.

Is this a hint to new physics?
More observational data, and new specifically designed experiments may help answering this question in the near future. 
With the data presently at hand we cannot exclude the possibility that all the anomalous observations are the result of inadequate understanding or poor statistics.
However, we cannot discard the \textit{new physics option} either. 
In particular, the observed additional cooling could be due to a novel \textit{weakly interactive slim particle (WISP)} \cite{Jaeckel:2010ni}, responsible for efficiently carrying energy away from the stellar interior.
This possibility is certainly appealing and should be examined in detail.

The purpose of this work is to investigate quantitatively the new physics solution and to select the candidate(s) which would better satisfy all the observations. 
Given the very different properties of the stars which show anomalous cooling, the novel production mechanisms should have a quite peculiar dependence on temperature and density. 
This allows to select only a few new physics candidates.

As we shall see, axions, or more generally axion-like particles (ALPs), are the best available candidates. 
The axion~\cite{Weinberg:1977ma,Wilczek:1977pj} is a light pseudoscalar particle predicted by the most widely accepted solution of the strong CP problem~\cite{Peccei:1977hh,Peccei:1977ur} and a prominent dark matter candidate~\cite{Abbott:1982af,Dine:1982ah,Preskill:1982cy}.
At low energies, its most relevant interactions with photons and fermions are described by the Lagrangian terms
\begin{eqnarray}
 {\cal L}_{\rm int}=- \frac{1}{4}g_{a\gamma} \, a F_{\mu \nu} \tilde F^{\mu \nu} - \sum_{\rm fermions}g_{ai} \, a \overline \psi_i \gamma_5 \psi_i \,,
\label{eq:Lint}
\end{eqnarray}
where $ g_{a\gamma}= C_{a\gamma} \alpha/(2\pi f_a)$ and $ g_{ai}= C_{ai} m_i/f_a$, with $ C_\gamma$ and $ C_i $ model dependent parameters and $ f_a $ a phenomenological scale known as the Peccei-Quinn symmetry breaking scale.
Moreover, in the so called QCD axion models, mass and Peccei-Quinn scale are related as\footnote{This mass-coupling relation is not a mandatory requirement for the solution of the strong CP problem, though relaxing it may require non-minimal assumptions~\cite{Rubakov:1997vp,Berezhiani:2000gh,Gianfagna:2004je}.} $ (m_{a}/1\mbox{ eV})= 6\times 10^6\, {\rm GeV}/f_a $.

More general models of pseudoscalar particles, known as ALPs, 
which couple to photons and, possibly, to fermions but do not satisfy the above mass-coupling relation, emerge naturally in various extensions of the Standard Model though, in general, their existence is not related to the strong CP problem~\cite{Ringwald:2012hr}. 
If light enough, these ALPs have been invoked to solve various astrophysical puzzles, including the quest of the transparency of the universe to very high energy (TeV) gamma rays (see sec.~\ref{sec:ALP}).

All this certainly accounts for the enthusiasm in the axion/ALP solution. 
However, presently, there is not a coherent analysis which excludes, just on the basis of the various observations, other possible candidates.

Neutrino electromagnetic form factors, if large enough, could, for example, account for some of the observed anomalies. 
In particular, a neutrino magnetic moment could explain the observed brightness of the RG tip in the M5 cluster~\cite{Viaux:2013hca}.
However, as we will show,  it would be inadequate to explain the observed deviations of the number distribution of WD per luminosity bin, the so called WD luminosity function (WDLF), from predictions, and would not be helpful in explaining the anomalous behavior of He-burning stars. 
Analogously, a neutrino minicharge would not help solving the anomalies observed in very dense stars, such as RGs and WDs. 

Another prominent new physics candidate is the Hidden Photon (HP), \M{a spin one field associated with a hidden U(1) gauge group.
The most general effective Lagrangian of dimension 4 describing this field is~\cite{Okun:1982xi,Holdom:1985ag}} 
\begin{equation}
\label{HPmodel}
{\mathcal L}^{(4)}_{V}= -\frac{1}{4} {F}_{\mu\nu}^{2} -\frac{1}{4} {V}_{\mu\nu}^{2} -\frac{\chi}{2} {F}_{\mu\nu} {V}^{\mu\nu} +
\frac{m^{2}}{2}{V}_{\mu\nu}{V}^{\mu\nu} \, , 
\end{equation}
where $ F $ and $ V $ represent, respectively, the standard photon and the HP fields, $ m $ is the HP mass and $ \chi $ is a dimensionless coupling constant.

Being massive spin one particles, HPs have two transverse and one longitudinal modes. 
Contrarily to what happens for ALPs, ordinary photons can oscillate into HPs without the need of an external magnetic or electric field. These oscillations can contribute to HP production in the stellar medium and, ultimately, to a more efficient stellar cooling. 

\M{This~} HP scenario has not been  considered so far in the interpretations of the cooling anomalies, perhaps since only recently a more clear description of the HP production rate became available~\cite{Redondo:2008aa,An:2013yfc,Redondo:2013lna,An:2014twa,Redondo:2015iea}.
As we will show, for a particular combination of mass and coupling HP could actually explain some of the cooling anomalies.
However, most of the allowed region in the parameter space hinted by the observations is excluded by other arguments.

What discussed above applies to the simplest \M{case of massive~} HP model \eqref{HPmodel}, in which the HP acquires its mass through the St\"uckelberg mechanism. Another possibility is that an extra (hidden) complex scalar charged under the hidden U(1) \M{(sometimes called Hidden Higgs, but here denoted as hidden scalar) } exists and takes a vacuum expectation value. In this case, the hidden \M{scalar }  behaves as a minicharged particle, due to the kinetic mixing~\cite{Holdom:1985ag}, and can be produced in stars. If the hidden scalar is light, the phenomenology would be essentially equivalent to that of any other minicharged particle~\cite{Ahlers:2008qc} and it could dominate the stellar losses~\cite{An:2013yfc}. If it is very massive, it decouples and one recovers the St\"uckelberg limit.  

\M{If the new vector is massless, the term proportional to $ \chi $, which describes a kinetic mixing between the standard and hidden photons, 
can be rotated away giving no observable consequences~\cite{Holdom:1985ag}. 
In this scenario, the most relevant interactions between the (massless) HP and the standard model fields emerge from dimension 6 operators~\cite{Dobrescu:2004wz}.
In particular, we should expect an interaction term with ordinary electrons $ e_{L,R} $ by virtue of an induced electron magnetic moment 
\begin{equation}
\label{magneticHP}
{\mathcal L}^{(6)}_{V}\ni \frac{1}{\Lambda^2} \bar l_L \sigma^{\mu\nu} H e_R V_{\mu\nu}+ h.c. \to 
\frac{v}{\Lambda^2} \bar e_L \sigma^{\mu\nu} e_R V_{\mu\nu} + h.c.  
\end{equation} 
where $v=246$ GeV is the Higgs vev~\cite{Dobrescu:2004wz}. Note that $SU(2)_L$ gauge invariance requires the presence of the Higgs vev and thus the dimension 5 operator at low energies emerges from dimension 6 and thus requires a suppression $\Lambda^2$ from the scale of new physics. 
At low energies where electrons are non-relativistic, the vector couples to the electron spin, just like  an ALP interacting with fermions~\cite{Hoffmann:1987et,Raffelt:1996wa}, so we will cover both cases together.}

In this paper we will analyze the available observational data in order to quantify the hints to new physics. 
We will proceed in the following way:
In section~\ref{sec:Cooling anomalies} we review the status of the cooling anomalies.
In section~\ref{sec:Model independent analysis of the WDLF} we analyze the WDLF and derive some hints from general emission rates which are expressed as a power-law of the temperature, in a model independent way. 
In section~\ref{sec:The case of a neutrino magnetic moment}, we begin a discuss of the most common new-physics candidates, starting with the neutrino magnetic moment.
In section~\ref{sec:MCP} we consider the case of minicharged particles and in section~\ref{sec:HP} the case of \M{massive } hidden photons.
The axion/ALP \M{and massless HP } cases are studied in detail in section~\ref{sec:ALP}, where we also consider the global significance of the combined hints from WD and RG on the axion-electron coupling.
Finally, in section~\ref{sec:Conclusion} we give our discussion and conclusion.

\section{Cooling anomalies}
\label{sec:Cooling anomalies}

Deviation from the standard cooling theory have been observed for over 20 years in various stellar systems.
The current observational status is summarized in Fig.~\ref{fig:WisePlot}, which shows the missing energy loss $ \Delta L $, normalized over a reference luminosity, $ L_{\rm st} $, for all cases where a statistical analysis of the uncertainties is available. 
The hinted anomalous luminosities, with their $ 1\,\sigma $ uncertainties, were calculated from the errors provided in the original references (see caption to Fig.~\ref{fig:WisePlot}). 

The original hint to a cooling anomaly was derived from the measurement~\cite{KeplerEtAl}, in 1991, of the period decrease $ \dot P /P $ of the star G117-B15A, a spectral type DA pulsating WD, a class known as DAV or ZZ Ceti. 
The reported measured period rate of change for the 215 s mode was $ \dot P =(12.0\pm 3.5)\times 10^{-15} $ s s$ ^{-1} $, substantially larger than what expected from standard pulsation theory $  \dot P =(2-6)\times 10^{-15} $ s s$ ^{-1} $ (see, e.g., \cite{Isern:1992gia}).

Since $ \dot P /P $ is practically proportional to the cooling rate $ \dot T /T $, it appears that this specific WD was cooling substantially faster than expected.
The anomalous energy loss, $ L_x $, can be estimated as~\cite{Isern:1992gia}
\begin{eqnarray}\label{Eq:Isern92_approx}
\dfrac{L_x}{L_{\rm st}}\simeq \dfrac{\dot P_{\rm obs}}{\dot P_{\rm th}}-1\,,
\end{eqnarray}
where $ L_{\rm st} $ is the standard energy loss.
Accordingly, even in the most optimistic hypothesis, one would find $ L_x\gtrsim L_{\rm st} $ from the results reported in~\cite{KeplerEtAl}.\footnote{Notice that Eq.~(\ref{Eq:Isern92_approx}) is strictly valid only when $ L_x\ll L_{\rm st} $, so that the unperturbed model can be used to estimate the energy loss.
However, numerical studies show that the above estimates are reasonable even for $ L_x\sim $ a few $ L_{\rm st} $.}

After several years of improved modeling and observations, G117-B15A still shows hints to exotic cooling. 
Additionally, two more WD variables have shown a similar anomalous behavior:
R548, a DA variable very similar to G117-B15A, and PG 1351+489, a DB variable (DBV) with quite different properties. 
The latest results are shown in table~\ref{tab:Pdot}.
\begin{table}[h]
\begin{center}
	\begin{tabular}{  l  l l l p{5cm}}
		\hline\hline
		 WD  			& class	&  $ \dPo $[s/s]  &  $ \dPt $[s/s]  \\ \hline
		G117 - B15A 	&DA 	&  $ (4.19 \pm 0.73)\times 10^{-15} $ 	&  $ (1.25 \pm 0.09)\times 10^{-15} $ \\ 
		R548 		 	&DA 	&  $ (3.33 \pm 1.1)\times 10^{-15} $ 	&  $ (1.1 \pm 0.09)\times 10^{-15} $ \\ 
		PG 1351+489		&DB 	& $ (2.0 \pm 0.9)\times 10^{-13} $ 		& $ (0.81 \pm 0.5)\times 10^{-13} $ \\ \hline 
	\end{tabular}
   \caption{Results for $\dot P $ for G117 - B15A~\cite{Corsico:2012ki},
   R548~\cite{Corsico:2012sh}, 
   and PG 1351+489~\cite{Corsico:2014mpa}. }
    \label{tab:Pdot}   
\end{center}
\end{table}

Besides this, various studies of the WD luminosity function (WDLF), which represents the WD number density per brightness interval, also seem to indicate a preference for an additional cooling channel. This has been quantified in the literature in terms of cooling by axion electron bremsstrahlung, pointing to  an axion-electron coupling 
$ g_{ae}\simeq (1.4\pm 0.3) \times 10^{-13} $ (at 1$ \,\sigma $)~\cite{Bertolami:2014wua}, while showing no improvement in the assumption of a neutrino magnetic moment~\cite{Bertolami:2014noa}.
In Fig.~\ref{fig:WisePlot}, we show the additional cooling deduced by the hinted results for the axion electron coupling in~\cite{Bertolami:2014wua}. 
Notice that $ \Delta L $, in the case of the WDLF, depends on the particular WD luminosity.
In Fig.~\ref{fig:WisePlot} we show the hinted anomalous cooling for an example of WD of intermediate brightness, $ M_{\rm Bol}\sim 9 $, where the bolometric magnitude $ M_{\rm Bol} $ is defined as $ M_{\rm Bol}=4.75 -2.5\log(L/L_{\odot})$, with $ L_{\odot} $ the solar luminosity.  

Further hints to anomalous energy loss in stars emerged in the recent analysis of the Red Giant Branch (RGB) stars in~\cite{Viaux:2013hca,Viaux:2013lha}.
The studies showed a brighter than expected tip of the RGB in the M5 globular cluster, indicating a somewhat over-efficient cooling during the evolutionary phase preceding the helium flash.
The anomalous brightness, $ \Delta M_{I,{\rm tip}}\simeq 0.2 $ mag in absolute $ I- $band magnitude, is within the calculated observational and modeling errors, which include uncertainties in the mass loss, treatment of convection, equation of state and cluster distance.
However, as evident from Fig~\ref{fig:WisePlot}, the error budget seems to just barely compensate for the difference between observed and expected brightness.
A reduction of the uncertainties,  particularly a better determination of the cluster distance, which may become possible with the GAIA mission, will certainly help clarifying the physical significance of this discrepancy. 

Finally, a recent analysis~\cite{Ayala:2014pea} showed a mild disagreement 
between the observed and the expected $ R $-parameter, 
$R= {N_{\rm HB}}/{N_{\rm RGB}}$, 
which compares the numbers of stars in the HB  ($N_{\rm HB}$) and in the upper portion of the RGB
($N_{\rm RGB}$).
More specifically, the observed value, $ R=1.39\pm 0.03 $ is somewhat smaller than the expected one 
$ 1.44\leq R\leq 1.50$. 
This indicates a surplus of RGB with respect to the numerical prediction and suggests that HB stars are cooling more efficiently, and therefore are less numerous, than expected. 

Besides what shown in Fig.~\ref{fig:WisePlot}, additional deviations from the standard cooling theory have been observed also in core He-burning stars of intermediate mass ($ M\sim 10M_\odot $). 
The problem, in this case, is that numerical simulations predict a larger number ratio of blue over red supergiants (B/R), with respect to what is actually observed~\cite{Skillman:2002aa,McQuinn:2011bb}. 
The predicted number would be lowered (alleviating or, perhaps, solving the B/R problem) in the hypothesis of an additional cooling channel efficient in the stellar core (but not in the H-burning shell)~\cite{Lauterborn:1971nva,Friedland:2012hj,Carosi:2013rla,Giannotti:2014cpa}. 
An exact prediction of the required additional cooling is, however, presently unavailable.

Finally, x-ray observations of the surface temperature of a neutron star in Cassiopeia A~\cite{Ho:2009mm,Heinke:2010cr,Shternin:2010qi} showed a cooling rate considerably faster than expected.
The effect seems to indicate the need for an additional energy loss roughly equal to the standard one. 
This was interpreted in terms of an axion-nucleon coupling  
$ g_{an}\sim 4\times 10^{-10} $~\cite{Leinson:2014ioa} or as due to a phase transition of the neutron condensate into a multicomponent state~\cite{Leinson:2014cja}.

The most remarkable aspect of all these results is that the hints seem to overwhelmingly point toward a positive $ \Delta L $, roughly of the same order of the standard luminosity, as clear from Fig.~\ref{fig:WisePlot}, indicating a systematic problem common to all the studied stellar systems.
The current inability  to reconcile these \textit{relatively large} anomalous observations with models shows the limits of our understanding of stellar evolution, either in the theoretical modeling or in the observations.
The current data is not sufficient to provide a clear solution to this problem. 
A new physics option is certainly appealing but should be critically investigated.
In the following sections we will show that, among the new physics candidates, ALPs provide an ideal solution. 
Interestingly, the ALP hinted region of the parameter space is accessible to the new generation ALP detectors, in particular ALPS II~\cite{Bahre:2013ywa} and the International Axion Observatory (IAXO)~\cite{Irastorza:2011gs,Armengaud:2014gea,Vogel:2015yka}.

\section{Model independent analysis of the WDLF}
\label{sec:Model independent analysis of the WDLF}

The analysis of the  WDLF can reveal interesting insights in new physics. 
Since WDs are (practically) only loosing energy, during their evolution they become less and less luminous.
Therefore, the number distribution of WD per luminosity bin provides information about the WD evolution and so about the cooling efficiency.

For a WD of fixed mass and chemical composition, the stellar luminosity is, with a good approximation, a function of the (central) temperature only. 
Therefore, an accurate analysis of the observed and expected WDLF may reveal not only the presence of an additional cooling channel, but also of its temperature dependence.

Some recent analyses of the WDLF at intermediate luminosities seem to indicate the need for additional cooling to properly fit the expected distribution of WDs .
However, the situation is far from clear since different observations seem to be in disagreement even within their reported errors, especially in the hottest bins, $ M_{\rm bol}\lesssim 7 $ (see~\cite{Bertolami:2014wua} and reference therein).

The WDLF has been used to test new physics hypothesis, in particular, the axion and the anomalous magnetic moment. 
Recent analyses showed a considerable improvement of the fitting when an ALP coupled to electrons is included as an additional cooling channel~\cite{Bertolami:2014wua}, hinting to an ALP-electron coupling $ g_{ae}=1.4^{+0.28}_{-0.35} \times 10^{-13}$ (at $ 1\,\sigma $), while a neutrino magnetic moment has shown negligible effects~\cite{Bertolami:2014noa}.

A more recent study of the hot part of the WDLF~\cite{Hansen:2015lqa} showed no significant improvement of the fit, even with the addition of an ALP coupled to electrons.
The hot section, however, has larger errors and the effect of axions may be hidden by the standard neutrino cooling there, since they have a production rate steeper in temperature. 
Further analysis is perhaps necessary to clarify this issue and we will not consider it further in this paper.

For the colder WDs, $ 7.75\lesssim M_{\rm bol}\lesssim 14.25 $, the neutrino cooling can be ignored and the particular shape of the WDLF can be described by a simple power law~\cite{Mestel,Shapiro:1983du,Raffelt:1996wa}:
\begin{eqnarray}\label{Eq:Lgamma}
L_\gamma=C_\gamma \Lsun \, T_7^{3.5}\,,
\end{eqnarray}
where $ C_\gamma $ is a dimensionless constant, and $ T_7 $ is the central temperature in units of $ 10^7 $K. 
In this case, the expected number of WD per luminosity bin, $ \Nexp $, can be approximated by $ \Napp $ given by
$ \log  \Napp =\frac27\, M_{\rm bol}+b  $, where $ b $ is a function of the WD characteristics.\footnote{See, e.g.,~\cite{Raffelt:1996wa}, sec. 2.2.2.}
Comparing the exact form of $ \Nexp $ from~\cite{Bertolami:2014noa} with Eq.~(\ref{Eq:Lgamma}) and using $ M=0.6\Msun $ we find the best fit value 
$ C_\gamma=8.5\times10^{-4}$. 

\begin{figure}[t]
\centering
\includegraphics[width=0.5\linewidth]{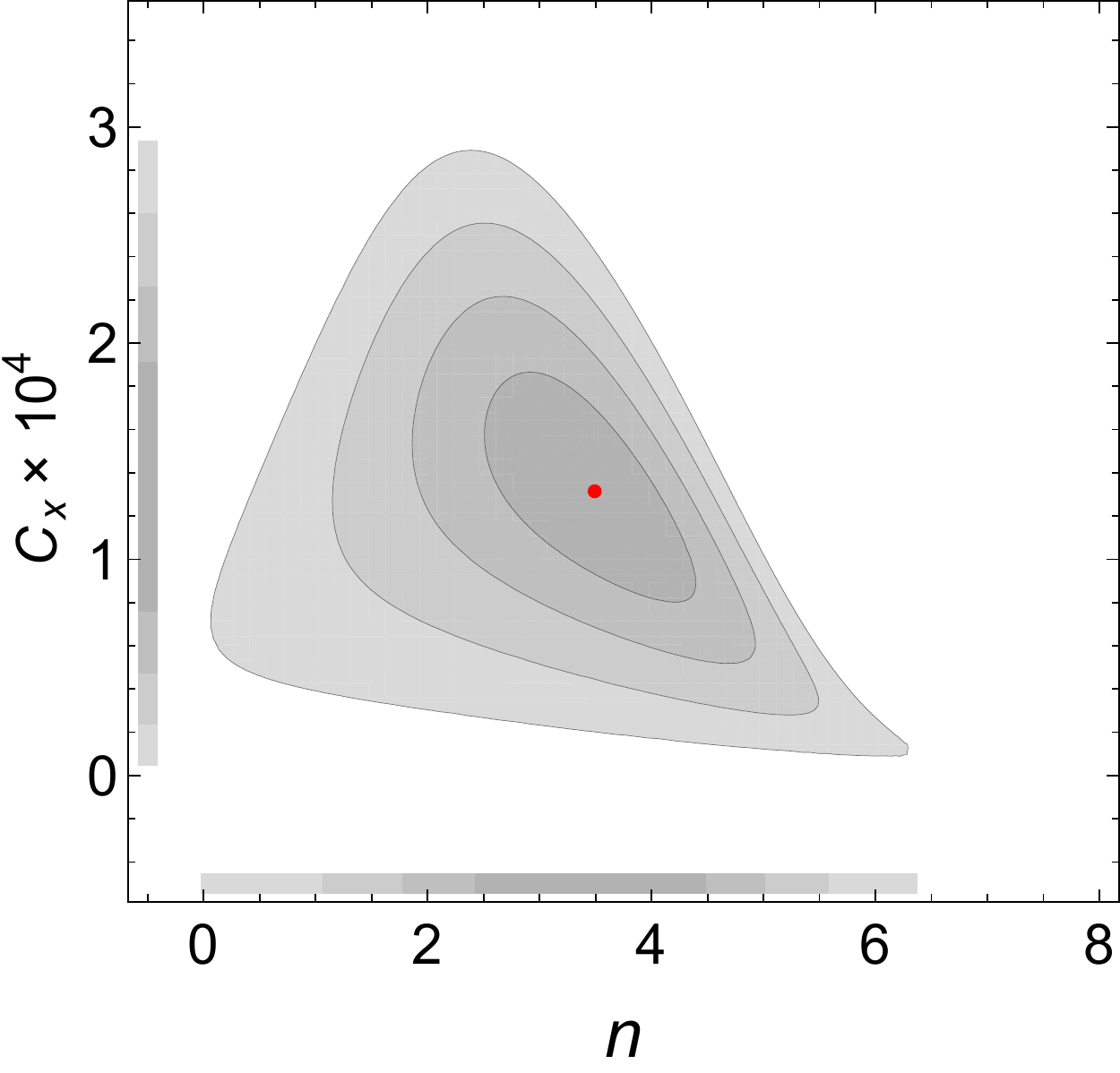}
\caption{Contours of the 1, 2, 3 and 4$ \,\sigma $ C.L. in the $ n $ and $C_x $ parameter space, defined in Eq.~(\ref{eq:Lx}).
The red dot refers to the best fit value: $ n_{\rm best}= 3.49 $, $ C_{x,{\rm best}}= 1.31\times 10^{-4} $.
The analysis is based on the data in~\cite{Bertolami:2014noa} for $ 7.75\lesssim M_{\rm bol}\lesssim 14.25 $.
}
\label{fig:WDLF_contour}
\end{figure}

Adopting the approximation in Eq.~\eqref{Eq:Lgamma}, allows a more general \textit{model-independent} analysis of exotic cooling rates
of the form
\begin{eqnarray}\label{eq:Lx}
L_x=C_x\,\Lsun \, T_7^n\,.
\end{eqnarray}
Let us define 
\begin{eqnarray}\label{eq:q_def}
q=\left( \Napp-N_{\rm obs}\right)/\Napp \simeq L_x/L_\gamma\,,
\end{eqnarray}
where the last equality is valid when $ L_x \ll L_\gamma $. 
Notice that $ q $ is a function of $ C_x $ and $ n $. 
A comparison of the measured $ q $ with the one calculated in terms of $ C_x $ and $ n $ can therefore provide insights about the two parameters characterizing the anomalous rate~(\ref{eq:Lx}).

We performed a $ \chi^2 $-test with our $ \chi^2 $ variable computed as 
\begin{eqnarray}
\chi^2(C_x,n)=\sum \left(\dfrac{q(C_x,n)^2-q}{\delta q}\right)^2\,,
\end{eqnarray}
with $ q $ calculated directly from $ \Napp $ and $ N_{\rm obs} $, and $ q(C_x,n) $ as a function of $ C_x $ and $ n $. 
The data for the analysis, including the uncertainties in $ N_{\rm obs} $, are taken from ref.~\cite{Bertolami:2014noa}. 
The errors in $ \Napp $ have been calculated, conservatively, as $ \delta\Napp=|\Nexp-\Napp| $.

With this hypothesis, we find $ \chi^2_{\rm min}=3.4 $ for $ n= 3.5 $ and $ C_{x}= 1.3\times 10^{-4} $.
The contours for the hinted parameters at 1 to 4$ \,\sigma $ are shown in Fig.~\ref{fig:WDLF_contour}.
The fit is excellent. 
We have used 14 data points for a two-parameter fit. 
So, $ N_{\rm d.o.f.}=12 $ and  $ \chi^2/N_{\rm d.o.f.}=0.28 $.
Notice also that, taking $ T_7\sim 1 $, the values of $ C_x $ show that indeed $ L_x\ll L_\gamma $, as we had originally assumed.

In general, our analysis indicates a clear preference for additional cooling, at about $ 4 \,\sigma $.
This is a higher confidence level than what is derived from the analysis in Ref.~\cite{Bertolami:2014wua}, which used a wider range of WD magnitudes and assumed the cooling to be induced by axions only (which corresponds to fixing $ n=4 $).

Although our procedure is necessarily approximate and cannot be extended to the hot part of the WDLF, it does show a preference for additional cooling and, as we shall see, in particular for ALPs or HPs, among the most common new-physics candidates. 

\section{The case of a neutrino magnetic moment}
\label{sec:The case of a neutrino magnetic moment}

Among the possible new physics candidates the neutrino magnetic moment is certainly one of the most widely studied. 
If equipped with a large (with respect to the standard model prediction) magnetic moment, neutrinos could be efficiently produced in a dense environment through an electromagnetically induced plasmon decay $ \gamma\to \nu\bar \nu $~\cite{Haft:1993jt}.

The most current bound on the neutrino magnetic moment is derived from the analysis of RG branch stars in M5~\cite{Viaux:2013lha,Viaux:2013hca}. 
The reported constraint is 
$ \mu_{\nu}\leq 2.6 (4.5) \times 10^{-12} \mu_{\rm B} $, at 68\% (95\%) C.L., where $ \mu_{\rm B} $ is the Bohr magneton.  
In the same analysis, it was also pointed out that a small neutrino magnetic moment, $ \mu_\nu \simeq 1.8 \times 10^{-12} \mu_{\rm B}$, could actually improve the agreement between the observed and predicted brightness of the RG
branch tip in M5 discussed in sec.~\ref{sec:Cooling anomalies}.

The recent asteroseismology analysis~\cite{Corsico:2014mpa} of PG 1351+489, a  WD variable of spectral type DB, hinted to a somewhat larger value, $ \mu_\nu\simeq 5\times 10^{-12} \mu_{\rm B} $.
This is a little above the 2$ \,\sigma $ bound from the RGB analysis. 
However, given the large uncertainties in the physics of the DBV, the two results are compatible and may indicate that a neutrino magnetic moment could provide a good solution at least for the anomalies observed in dense stars.

As we shall see, this is not the case. 
Allowing neutrinos to have a neutrino magnetic moment of a few $ 10^{-12} \mu_{\rm B} $ would not help explaining the observed WD behaviors.

As shown in~\cite{Haft:1993jt}, the electromagnetically induced plasmon emission rate has the same temperature dependence of the standard plasmon process, which, at the high density characterizing the WD interior, is very steep in $ T $.
More precisely, in the environment typical of the WD interior the neutrino emission rate is roughly proportional to $ T^8 $~\cite{Isern:2008nt}.
As shown in sec.~\ref{sec:Model independent analysis of the WDLF}, this temperature dependence 
is too steep to account for the observed behavior of the WDLF and is excluded at more than $ 4\,\sigma $.
This result is in agreement with the analysis in Ref.~\cite{Bertolami:2014noa} which showed no substantial improvement of the WDLF (in this case the analysis included also the hotter section) when a neutrino magnetic moment was added.

Additionally, again because of this steep temperature dependence, it would be impossible for a neutrino magnetic moment to explain simultaneously the pulsation problems observed in the DA and DB variables.
A rough estimate is enough to prove this statement.
The internal temperature in the DA stars is about a factor of 3.8 lower than the one in PG 1351 (compare Fig.~1 in~\cite{BischoffKim:2007ve} and Fig.~4 in~\cite{Corsico:2014mpa}). 
Hence, according to the temperature dependence of the production rate, the magnetic moment induced rate would be suppressed by a factor of roughly $ 4\times 10^4 $.
However, the hinted luminosity in the DA WDs is only about an order of magnitude smaller than in PG 1351.
Since the neutrino rate is proportional to $ \mu_\nu^2 $, one would need an enormous neutrino magnetic moment to explain the additional cooling observed in the DA variables, something about two orders of magnitude above the current upper limit.

Besides this, the magnetic moment induced plasmon decay rate (just like the standard one) is largely suppressed in low density environments.
In fact, because of the peculiar plasmon dispersion relation~\cite{Raffelt:1996wa}, a low plasma frequency kinematically restricts the ability for plasmon to decay into neutrinos. 
Hence, for any reasonable value of $ \mu_\nu $, the additional magnetic moment induced cooling would be negligible during the HB evolution~\cite{Raffelt:1996wa} or the He-burning stage of a massive star~\cite{Heger:2008er}, and would therefore be inadequate to explain the cooling excesses observed in those stars.

\section{Minicharged particles}
\label{sec:MCP}

Another widely discussed option for physics beyond the standard model is a minicharged particle (MCP).
A MCP is a fermion $ \psi $ coupled to ordinary photons $ A $ just like standard model fermions, however with a very small charge: ${\cal L}_{\rm int} = \epsilon \,e \, \bar \psi \gamma_\mu \psi A^\mu $, with $ e $ the electron charge and $ \epsilon \ll 1$ a phenomenological parameter~\cite{Davidson:2000hf}. \M{Additionally, one expects a dimension 5 magnetic operator and 4-fermion dimension 6 operators to appear at some level in the effective field theory where the new physics interacting with both the SM and the MCP has been integrated out. These terms are suppressed by a new energy scale $\Lambda$ which is expected to be above the TeV. This suppresses enormously the MCP production at the typical temperatures $\sim$keV of stellar interiors and introduces additional powers of the temperature in the energy loss production rates with respect to the minicharge. The discussion of these terms is parallel to that of the non-standard neutrino emission and, we argue, does not help fitting the anomalies. So we will concentrate on the minicharge henceforth.  }

Light fermions with a small charge could be produced through plasmon decay, just like ordinary neutrinos or neutrinos with an anomalous magnetic moment.
The emission rate would in this case have the same temperature dependence as these processes, but a less steep density dependence~\cite{Haft:1993jt}. 

Because of the weaker density dependence, an MCP could be more efficiently produced in low density stars such as HBs or the sun~\cite{Raffelt:1996wa}.
The most recent bound, $ \epsilon\leq 2\times 10^{-14}$ (95\% C.L.), is derived from a careful helioseismiological analysis of the sun~\cite{Vinyoles:2015khy}, and corresponds to the results previously obtained from the analysis of 
RG branch and HB stars~\cite{Davidson:2000hf}.

The MCP, however, offers an inadequate solution to the cooling anomalies. 
In particular, because of the steep temperature dependence we can draw the same conclusions about WDs derived for the neutrino magnetic moment. 
Namely, this new physics mechanism would not permit the simultaneous explanation of the DA and DB WD variables, nor improve the fit of the WDLF.

Similar conclusions can be drawn when a massive HP mediates, through kinetic mixing with the standard photon, interactions between the standard and a dark sector. In this scenario, the charged fields of the standard model become (mini)charged under the new Abelian gauge symmetry and this would enhance the stellar cooling efficiency. In this case, the temperature dependence of the loss-rate is expected to be $T^6$~\cite{Dreiner:2013tja}, still too steep to explain the WDLF anomaly.

\section{Hidden photons}
\label{sec:HP}

Hidden Photons have enjoyed a considerable interest revival recently~\cite{Redondo:2008aa,An:2013yfc,Redondo:2013lna,Dreiner:2013mua,Dreiner:2013tja,An:2014twa,Schwarz:2015lqa}. 
However, there is currently no study of the cooling excesses in terms of HPs. 
In this section we analyze separately the hints to the HP parameter space from our analysis of the cold section of the WDLF and from the other cooling anomalies. 

\subsection{Hints from the WD luminosity function}
\label{sec:Hints to HP from the WD luminosity function}

Here we apply the results of sec.~\ref{sec:Model independent analysis of the WDLF} to find for which parameters the emission of HPs from WDs could substantially improve the WDLF fit.

The analysis, in this case, is not quite as straightforward as for the other candidates discussed in this work.
In fact, the HP production rate, dominated by the emission of longitudinal modes, is, in general, not a power law of the temperature.
Additionally, the rate is a function of both the HP mass and coupling constant.  

In the case of low mass, the emission rate of longitudinal HPs is~\cite{Redondo:2013lna}
\begin{eqnarray}\label{eq:HP_rate}
\varepsilon\simeq\dfrac{\chi^2\, m^2\,\omega_{\rm pl}^3}{4\pi \rho\left(e^{\omega_{\rm pl}/T}-1\right)} \,,
\end{eqnarray}
where $ \omega_{\rm pl} $ is the plasma frequency of the WD in the location where the HP energy $ \omega $ satisfies the resonance condition 
$ \omega=\omega_{\rm pl} $.

In the center of a WD, $ \omega_{\rm pl}\simeq 30 $\,keV \M{while the temperature is a few keV. 
Therefore, the exponential factor $e^{\omega_{\rm pl}/T}$ in the denominator is huge and suppresses considerably the HP emission.  The HP production will then peak somewhat off center where the density has decreased sufficiently. 

The rate in Eq.~(\ref{eq:HP_rate}) can be approximated as a power law in the temperature around a relevant temperature $T_0\sim (2-3) $\,keV 
\begin{equation}
n = \frac{d \log \varepsilon}{d \log T} \simeq \frac{\omega_{\rm pl}}{T_0} .  
\end{equation}
A numerical analysis performed on a grid of WD models shows that the HP emission rate (for masses well below $ \omega_{\rm pl}$) follows a power law with $n\sim 3.7$ with surprising accuracy. This corresponds to a HP emission from a shell where the density has decreased by a factor of 10 or so from its central value.  
The numerical result shows that, according to our discussion in sec.~\ref{sec:Model independent analysis of the WDLF}, for particular combinations of coupling and mass HPs can indeed help the WDLF fit. } 

At small masses, since the emission rate is proportional to  
$(\chi m)^2 $,
we see that the hinted HP region should be a band in the $ \log \chi- \log m $ parameter space between two parallel lines of slope $-1$.
Our numerical calculation of the emission rate (without approximating the exponent and without assuming a small mass) confirms this behavior, as shown in the left panel of Fig.~\ref{fig:summary_HP}.

\subsection{Hints from WD pulsation, RG and HB stars}
\label{sec:Hints to HP from WD pulsation, RG and HB stars}

\begin{figure}[t]
	\centering
	\includegraphics[width=0.4\linewidth]{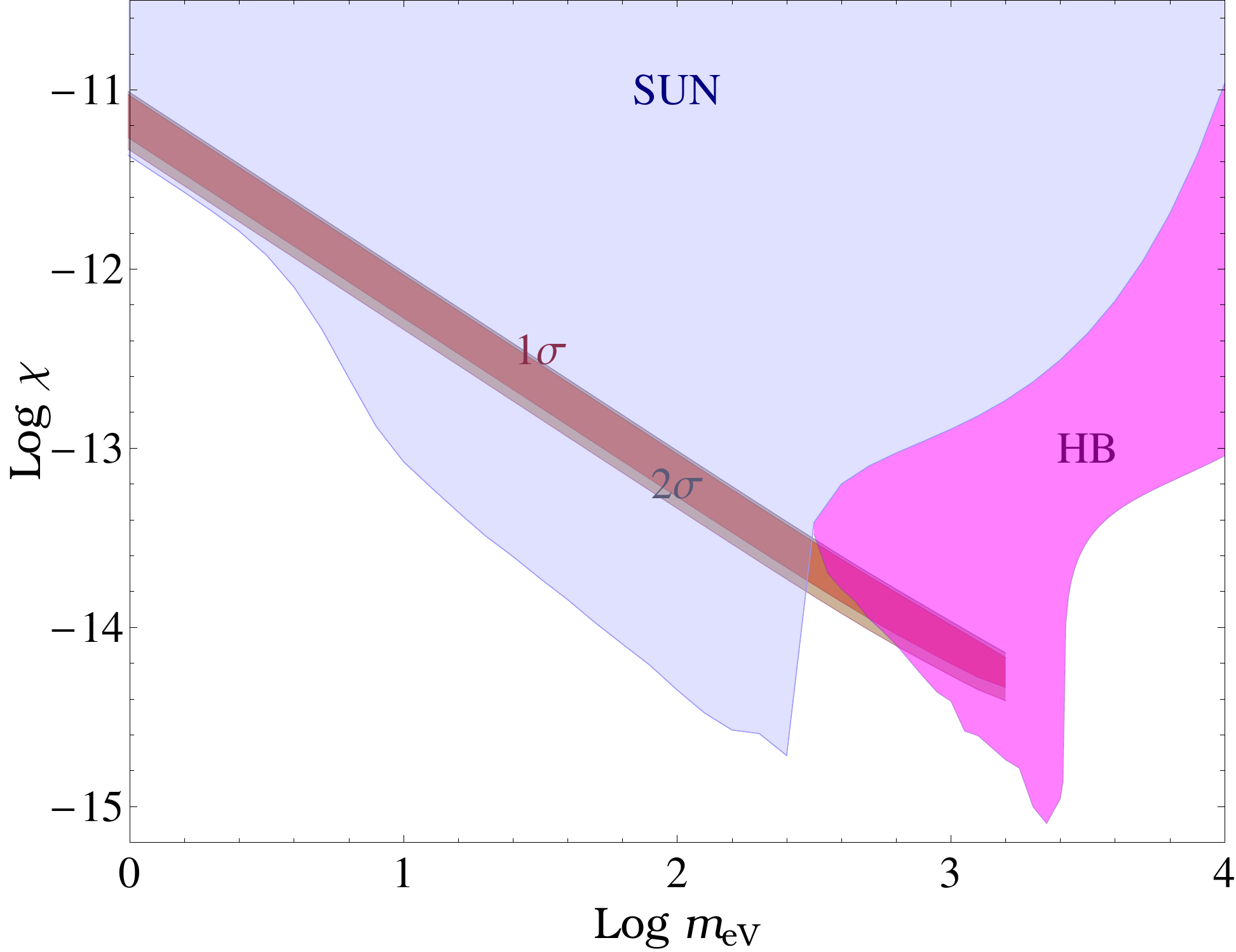}
	\hspace{1cm}
	\includegraphics[width=0.45\linewidth]{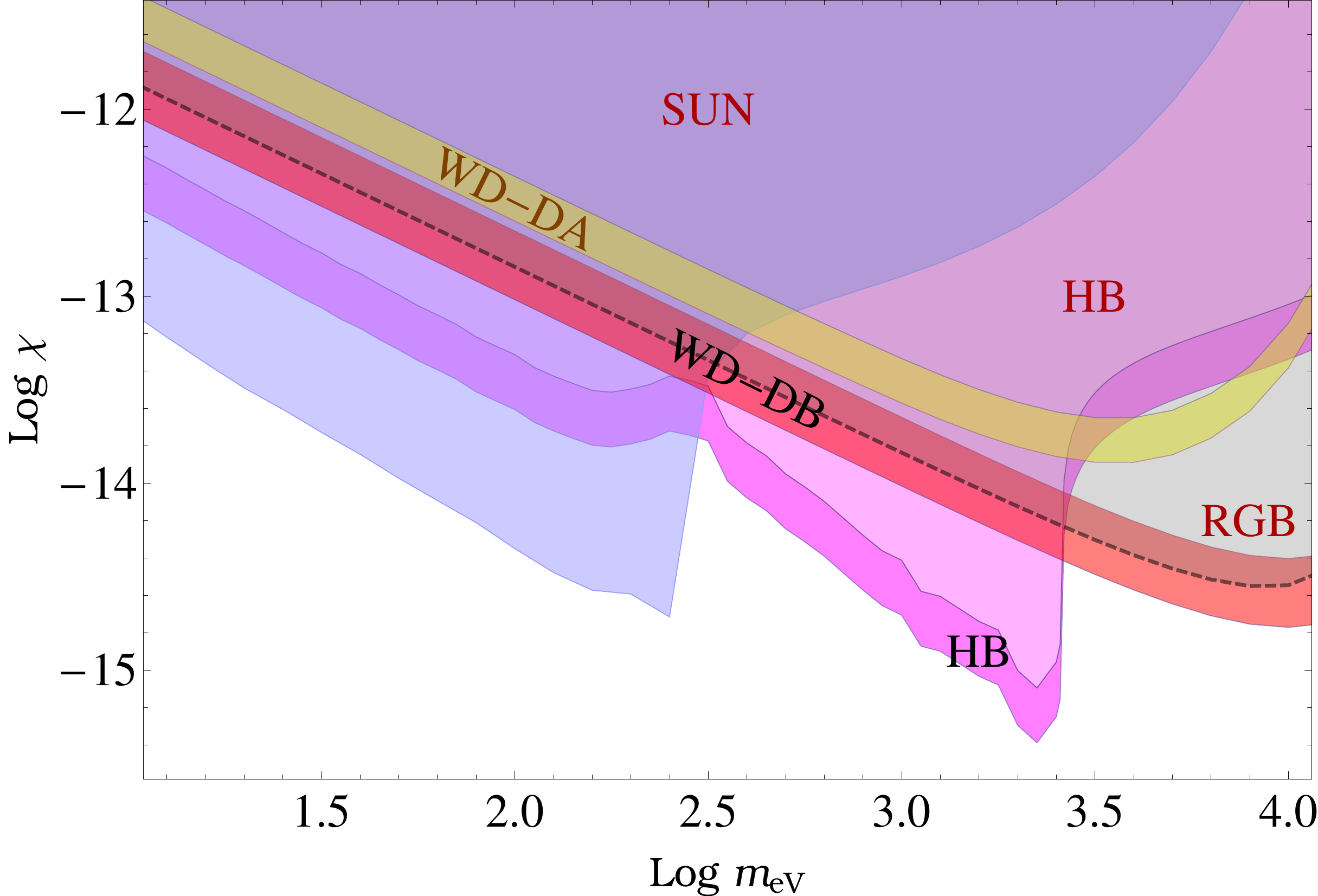}
	\caption{Summary of the regions in the HP parameter space hinted by the observed cooling anomalies. 
	\textit{Left panel:} Areas in the HP mass-coupling parameter space corresponding to the 1 and 2$ \,\sigma $ regions shown in Fig.~\ref{fig:WDLF_contour}.
		The numerical analysis was performed on a grid of WD models simulated with MESA~\cite{MESA}.
		The regions excluded by the sun and the HB stars are also shown.
		\textit{Right panel:}  Summary of the hints to HP from the WD pulsation (the yellow band represents the region hinted by the analysis of R548 while the orange band that hinted by PG 1351+489) and HB stars (dark magenta band labeled in black). 
		Superimposed are the constraints from the sun (blue area), HB (light magenta area), and RGB stars (gray area above the dashed line).
		For clarity, we have not included the region hinted by the WDLF, shown in the left panel.}
	\label{fig:summary_HP}
\end{figure}

As said, there is currently no study of the cooling anomalies in terms of HPs. 
In particular, there is no analysis with simulations which include HPs as a possible cooling channel.
We can, however, gain some insight on the hinted HP parameter space by comparing the anomalous luminosities, shown in Fig.~\ref{fig:WisePlot}, with the HP rates calculated from standard (unperturbed) stellar models.
The result of this analysis is shown in the right panel of Fig.~\ref{fig:summary_HP}.

The solar exclusion region is calculated by integrating the HP emission rate, both the transverse and longitudinal contributions, over the standard solar model~\cite{Serenelli:2009yc} and selecting only the parameter space such that $ L_x\leq 0.1 \Lsun $.

For the RGB and HB stars, we used typical models derived by means of the FUNS  (FUll Network Stellar evolution) code~\cite{Straniero:2005hc,Luciano:2013rya, straniero2014} (curtesy of Oscar Straniero).
Based on the discussion in~\cite{An:2014twa}, as dominant cooling channel we used the HP L-mode for RGB stars and the HP T-mode for HB stars. 
The HB hinted region (light magenta area) corresponds to the luminosity shown in Fig.~\ref{fig:WisePlot}, derived from the $ 1\,\sigma $ interval for $ g_{a\gamma} $ in~\cite{Ayala:2014pea}.
For the RGB, we are showing only the exclusion plot, calculated according to the results in the most recent analyses~\cite{Viaux:2013lha,Viaux:2013hca}.

In the case of the pulsating WDs, we compared the measured and expected period change as reported in~\cite{Corsico:2012sh} for R548 (which we chose as a representative for the DA class) and~\cite{Corsico:2014mpa} for PG 1351+489.
In both cases, we derived the exotic energy loss using Eq.~(\ref{Eq:Isern92_approx}).
This could be not completely accurate since the additional energy loss is quite large, especially in the case of R548.
The hinted HP parameter space (mass and coupling) is derived using MESA simulations of WD models with characteristics similar to those used in the refereces above.
The results are shown in the yellow (WD-DA) and orange (WD-DB) bands in Fig.~\ref{fig:summary_HP}.
In the case of the pulsating WD PG 1351+489, the $ 1\,\sigma $ interval could extend to include the standard case, where no additional cooling is necessary. 
So, the lower bound in the hinted band in Fig.~\ref{fig:summary_HP} is only indicative and could extend to lower couplings.

Although these results are based on several simplifications and the exact regions could be slightly modified, it is clear from the figure that a HP cannot easily accommodate the different hints, no matter its mass and coupling. 
A more detailed analysis, however, should require a full simulation of the stellar evolution including the HP cooling channel.

\M{We finally notice that the strong bounds on the HP vanish in models of a massless vector field with leading interaction terms emerging from dimension 6 operators~\cite{Hoffmann:1987et,Dobrescu:2004wz}. 
The massless HP production rate in WD and RGB has the same functional form as that of an ALP coupled to electrons, which is discussed in sec.~\ref{sec:ALP}.
 }

\section{Axion-like particles}
\label{sec:ALP}

ALPs have enjoyed a considerable attention lately, in relation to several astrophysical observations.

Besides being able to explain the stellar cooling anomalies, as we shall discuss further in this section, and being dark matter candidates~\cite{Arias:2012az}, 
 light ALPs have been invoked to explain other astrophysical anomalies, such as the seeming transparency of the universe to very high energy (TeV) gamma rays in the galactic and extragalactic medium~\cite{De Angelis:2007dy,Horns:2012fx,Meyer:2013pny} and some anomalous redshift-dependence of AGN gamma-ray spectra~\cite{Galanti:2015rda} (though this last hypothesis currently shows some conflict with the SN1987A bound on the axion-photon coupling~\cite{Payez:2014xsa}).
More recently, it was also pointed out that anomalous X-ray observations of the active Sun suggest an ALP-photon coupling of the same size hinted by the other analyses~\cite{Rusov:2015sqa}.  
Interestingly, the required couplings are not excluded by experiments nor by phenomenological considerations and are accessible to the new generation ALP detectors, in particular ALPS II~\cite{Bahre:2013ywa} and the International Axion Observatory (IAXO)~\cite{Irastorza:2011gs,Armengaud:2014gea,Vogel:2015yka}.

In this section we present the hints on the axion/ALP coupling from the cooling anomalies, based on the published literature. 
In sec.~\ref{sec:Hints to the ALP-photon coupling from HB stars} and ~\ref{sec:Hints to the ALP-nucleon coupling from neutron stars} we review, respectively, the hints on the ALP-photon coupling from helium burning stars and on the ALP-nucleons coupling from neutron stars.
In sec.~\ref{sec:Hints to the ALP-electron coupling from WD and RG} we propose a new global analysis for the hints on the ALP-electron coupling from all the combined anomalous observations. 

\subsection{Hints to the ALP-photon coupling from He-burning stars}
\label{sec:Hints to the ALP-photon coupling from HB stars}

The analysis of the $ R $-parameter in~\cite{Ayala:2014pea} showed a mild preference for additional cooling, as shown in Fig.~\ref{fig:WisePlot}.
This can be interpreted as a hint to an ALP, produced through Primakoff process in the HB core.
Such an analysis indicates
\begin{eqnarray}\label{Eq:1sigma}
g_{a \gamma} = 0.45_{-0.16}^{+0.12} \times 10^{-10}~{\rm GeV}^{-1} \,\  \,\ (68 \% \,\ \textrm{CL})
\,,\hspace{3ex} {\rm for}\ m_a\lesssim 30\, {\rm keV}  \, ,
\end{eqnarray}
for the ALP-photon coupling.  
At 2$ \,\sigma $ the result for the coupling is compatible with zero.
Remarkably, the hinted value for $g_{a \gamma}  $ is roughly the same of what required for the ALP explanation of the observed transparency hints -- the latter requiring however $m_a\lesssim 10^{-7}$\,eV --  and is in full reach of the next generation ALP detectors~\cite{Irastorza:2011gs,Bahre:2013ywa,Armengaud:2014gea,Vogel:2015yka}.

A recent revision by one of us (MG), which included errors from the most important nuclear reactions and a more careful determination of the $ R $-parameter, suggested a more conservative estimate~\cite{OscarPatras},  
\begin{equation}
g_{a \gamma} = (0.29\pm {0.18}) \times 10^{-10}~{\rm GeV}^{-1} \,\  \,\ (68 \% \,\ \textrm{CL})
\,,\hspace{3ex} {\rm for}\ m_a\lesssim 30\, {\rm keV}  \, .
\end{equation} 

Additionally, the observed smaller than predicted number ratio of blue over red supergiants in open clusters (see Ref.~\cite{McQuinn:2011bb} and references therein) could also point to an ALP-photon coupling~\cite{Friedland:2012hj,Carosi:2013rla,Giannotti:2014cpa}.
ALPs produced through the Primakoff process would be efficient in the He-burning stellar core, while would be irrelevant in the colder H-burning shell. 
Such a mechanism is known to accelerate the blue loop stage, reducing therefore the number of blue with respect to red supergiants~\cite{Lauterborn:1971nva}. 

The analysis in Ref.~\cite{Friedland:2012hj} indicated that an axion-photon coupling of a few $ 10^{-11}\, {\rm GeV}^{-1}$, in the same range as the one hinted by the HB anomaly, would reduce the number of expected blue stars, alleviating or perhaps solving the anomaly. 
However, in this case the uncertainties in the microphysics and in the observations are, essentially, unquantifiable. 

We finally notice that the emission of ALPs with a comparative coupling may affect the nucleosynthesis in massive stars, particularly the Neon yield~\cite{Aoyama:2015asa}. However, at the moment the predictions need quite some refinement and reliable observables are missing.

\subsection{Hints to the ALP-nucleon coupling from neutron stars}
\label{sec:Hints to the ALP-nucleon coupling from neutron stars}

Recently, x-ray observations of the surface temperature of a neutron star (NS) in the supernova remnant Cassiopeia A showed a cooling rate considerably faster than expected.
The effect may be interpreted in terms energy loss due to ALP bremsstrahlung on nucleons, requiring an ALP-nucleon coupling~\cite{Leinson:2014ioa} of the order of
\begin{eqnarray}
 g_{an}\sim 4\times 10^{-10} \,,
\end{eqnarray}
within the current limits from NS cooling~\cite{Keller:2012yr,Sedrakian:2015krq}.

A confidence level in this result is at the moment unavailable. 
Moreover, the uncertainties in the physics of neutron stars cooling make this only a marginal hint and the effect could have a different origin~\cite{Leinson:2014cja}.
If one insists on an ALP interpretation, it would be important to make sure that such a coupling does not shorten significantly the neutrino pulse of the SN1987A. It has been pointed out recently that this anomaly is compatible with the constraints and the WD and RG hints if the ALP is a QCD axion of the DFSZ type~\cite{redproc}.

\subsection{Hints to the ALP-electron coupling from WDs and RGs}
\label{sec:Hints to the ALP-electron coupling from WD and RG}

All the observed anomalies from WDs (both pulsation and luminosity function) and RGB stars can be interpreted as a hint to a possible ALP-electron coupling.
We shall argue that the same applies also to the HB star hint, discussed in sec.~\ref{sec:Hints to the ALP-photon coupling from HB stars},
which so far has  been studied only in relation to the ALP-photon coupling.

The dominant ALP production mechanism in a WD or RGB star is electron brems\-strah\-lung, $ e + X\to e+X+a $, where $ X $ is either an electron or a nucleus and $ a $ is the ALP. 
The emission rate for this process is proportional to $ T^4 $~\cite{Nakagawa:1987pga}, a power law of the temperature with an exponent remarkably close to our best fit value in the analysis of the WDLF in sec.~\ref{sec:Model independent analysis of the WDLF}.

In He-burning stars, the Compton production, $\gamma+e^-\to a+e^-$,  contributes as much as the bremsstrahlung processes. Using the HB models of~\cite{Dearborn:1989he} and the emission rates summarized in~\cite{Redondo:2013wwa} we obtain the HB axion luminosity 
$L_{g_{ae}}\simeq 1.33 \alpha_{26}L_\odot$, with $ \alpha_{26}=g_{ae}^2/4\pi $ in units of $ 10^{-26} $, and a reduction of the HB lifetime 
$\delta t_{\rm HB}/t_{\rm HB}\simeq -0.067 \alpha_{26}$, where we adopted a typical Helium fusion luminosity of $20L_\odot$. 

\begin{figure}[t]
\begin{center}
	\centering
\includegraphics[width=6cm]{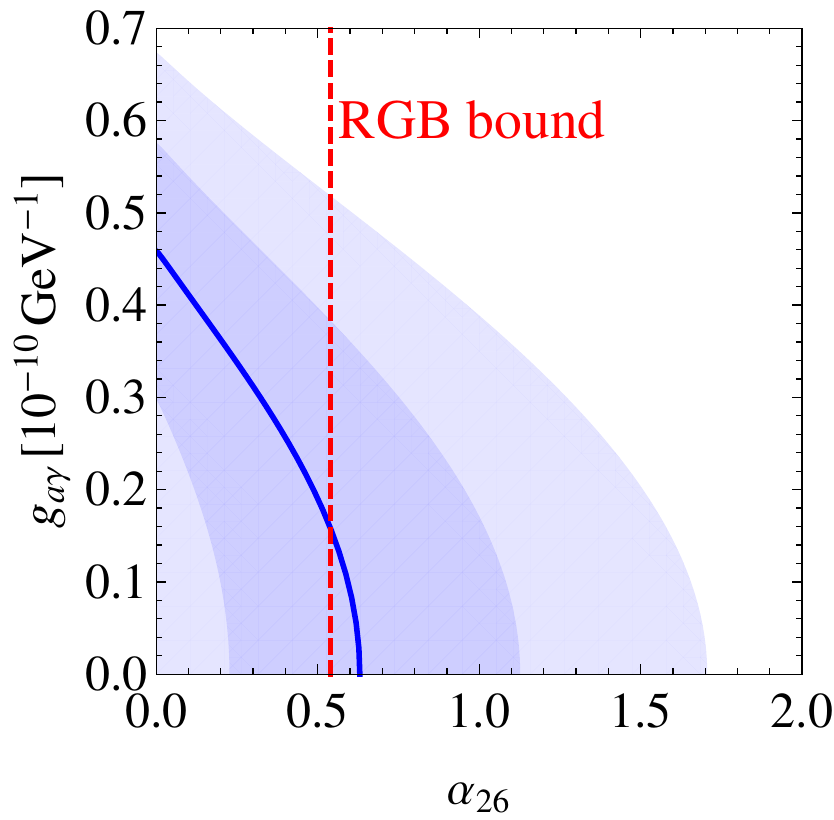}
	\hspace{1cm}
\includegraphics[width=6cm]{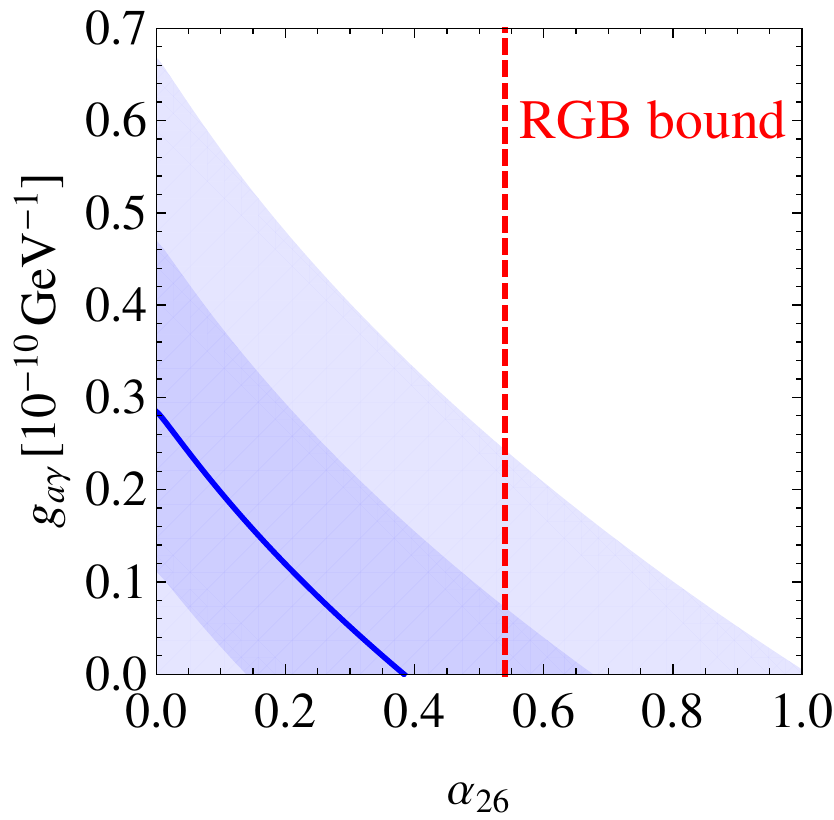}
\caption{Hinted 1 and $ 2\, \sigma $ regions from the analysis of the $ R $-parameter in sec.~\ref{sec:Hints to the ALP-electron coupling from WD and RG}.
The current bound from the RGB analysis in M5~\cite{Viaux:2013lha} is also shown.}
\label{fig:HBgae}
\end{center}
\end{figure}

The ALP-electron coupling has also an indirect effect on the $ R $-parameter. 
The energy loss during the RGB phase leads to a larger core mass. 
From the recent simulations~\cite{Viaux:2013hca,Viaux:2013lha} of the RGs of the cluster M5 we found the analytical approximation, valid for 
$ \alpha_{26}\lesssim 2.8 $,
\begin{eqnarray}
\delta {\cal M}_c= 0.024 M_\odot \left(\sqrt{4\pi\alpha_{26}+(1.23)^2}-1.23-0.921 \,\alpha_{26}^{0.75}\right) \,.
\end{eqnarray} 
With both these additions, the ratio of HB to RG stars presented in~\cite{Ayala:2014pea} is updated to 
\begin{equation}
R= 6.26 Y - 0.12 -0.14 g^2_{10} -1.61 \delta {\cal M}_c -0.067 \alpha_{26} ,  
\end{equation}
where $ \delta M_c $ is in solar mass units, $ Y $ is the primordial He abundance and $ g_{10}=g_{a\gamma}\times 10^{10} $GeV. 
The factor $\frac{d R}{d {\delta {\cal M}}_c}=-0.7\times \ln(10)=-1.61 $ was derived in \cite{buzzoniR,Raffelt:1989xu}. Repeating the analysis of \cite{Ayala:2014pea} but now with two possible energy loss channels, we find the $1\sigma,2\sigma$ intervals depicted in the left panel of Fig.~\ref{fig:HBgae}.

Notice that a value of $ \alpha_{26}= 0.63^{+0.5}_{-0.4}$ could account for the anomaly even in the absence of the ALP-photon channel. 
Although -- within the error -- this possibility cannot be excluded, the required coupling is slightly above the most current bound $ \alpha_{26}< 0.54 $~\cite{Viaux:2013lha}. Additionally, a comparison with the other hints, along the line discussed below, shows that the assumption that the HB hint could be explained solely by an ALP-electron coupling is quite disfavored ($ \chi^2_{\rm min} =20$, for 4 d.o.f.).

The result changes slightly when we adopt the new analysis in~\cite{OscarPatras}.
In this case, we find a somewhat different prediction for the $ R $-parameter dependence on the ALP-photon coupling,
\begin{eqnarray}
R= 7.33 Y + 0.02 -0.095 \sqrt{21.86+21.08 g_{10}}-1.61 \delta {\cal M}_c -0.067 \alpha_{26} ,
\end{eqnarray}
which leads to the 1 and $2\,\sigma$ intervals shown in the right panel of Fig.~\ref{fig:HBgae}.

In this case, the suggested value for the ALP-electron coupling necessary to explain the HB hint without requiring also an additional ALP-photon coupling is lowered to $ \alpha_{26}= 0.38^{+0.3}_{-0.24}$, within the currently allowed range and more in agreement with the other hints, as discussed below.

The hinted values from the different analyses are shown in Tab.~\ref{tab:g13_squared}.
The last row refers to the hint from the $ R $-parameter discussed above (the HB hint) but with the assumption that the ALP interacts only with electrons. 
Notice that an additional ALP-photon coupling would change this hinted value, as shown in Fig.~\ref{fig:HBgae}, or even eliminate it, as discussed in sec.~\ref{sec:Hints to the ALP-electron coupling from WD and RG}.
The other hints, on the other hand, would be essentially unaffected by an ALP-photon coupling below the allowed bounds.
Therefore, we treat these hints on a different footing and, in presenting the analysis of the data in the table, we will show separately the cases where we include or not the HB hint.

\begin{table}[h]
	\begin{center}
		\begin{tabular}{ l   c   c  p{5cm} |}
			\hline \hline
							&  $ \alpha_{26} $ 			& references \\ \hline 
			G117 - B15A 	& $ 1.87\pm 0.53 ~ $  		& \cite{Corsico:2012ki}  \\ 
			R548 			& $ 1.82\pm 1.03 ~ $   		& \cite{Corsico:2012sh} \\ 
			PG 1351+489		& $ 0.52\pm 0.67~ $  		& \cite{Corsico:2014mpa} \\
			WDLF  			& $ 0.156\pm 0.068 $   		& \cite{Bertolami:2014wua} \\ 
			RG 				& $ 0.26\pm 0.28 $   		& \cite{Viaux:2013lha}\\ 
			HB				& $ 0.38\pm 0.3 $   		& this work \\ \hline
		\end{tabular}
		\caption{Results for $\alpha_{26} $ at $ 1\, \sigma $ from WD pulsation, WDLF and RGB stars. In the case of the WDLF analysis, the $ 1\,\sigma $ interval is extracted from the inset in fig.~6 in~\cite{Bertolami:2014wua}.}
		\label{tab:g13_squared}   
	\end{center}
\end{table}

All the observations hint to a positive coupling square (therefore to an additional energy loss) though in two cases the $ 1\,\sigma $ intervals are compatible with the standard model solution or even with an additional energy source. 
In the case of 	PG 1351+489, there is no study which considered the axion and we derived the $ 1\,\sigma $ limits on $ \alpha_{26} $ shown in Tab.~\ref{tab:g13_squared} using the approximate formula~(\ref{Eq:Isern92_approx}).

What can we learn from the results in Tab.~\ref{tab:g13_squared}?

First of all, it is noticeable that the results from the two DA WD, G117 - B15A and R548, hint to a much larger coupling constant than all other observations. 
This may be due to the hypothesis that the particular oscillating mode examined (in both cases, the one with period about 215 s) is trapped in the envelope (see, e.g., \cite{Corsico:2012ki}).
This assumption may be incorrect (see, e.g. discussion in~\cite{Bertolami:2014wua}) although currently there is no certainty and this is still subject to speculations.

Relaxing the trapping hypothesis, the period change can be
estimated as $ \dot P\simeq 3.9\times 10^{-15} $s/s~\cite{Corsico:2012ki}, now within 1$ \,\sigma $ of the measured one, though still smaller.
Interestingly, in this case we would expect an additional luminosity corresponding to $ g_{ae}\simeq  10^{-13} $, quite closer to what predicted by RG and WDLF. 
More accurately, from Eq.~(\ref{Eq:Isern92_approx}) we find 
$ L_x=(0.074\pm 0.19) L_{\rm st}$, with $ L_{\rm st}=1.2\times 10^{31} $erg/s~\cite{Corsico:2012ki}.
So, using for example Eq.~(6) in reference~\cite{BischoffKim:2007ve} for the ALP emission, we find $ \alpha_{26}= 0.079 \pm 0.201 $.

\begin{figure}[t]
	\centering
	\includegraphics[width=0.35\linewidth]{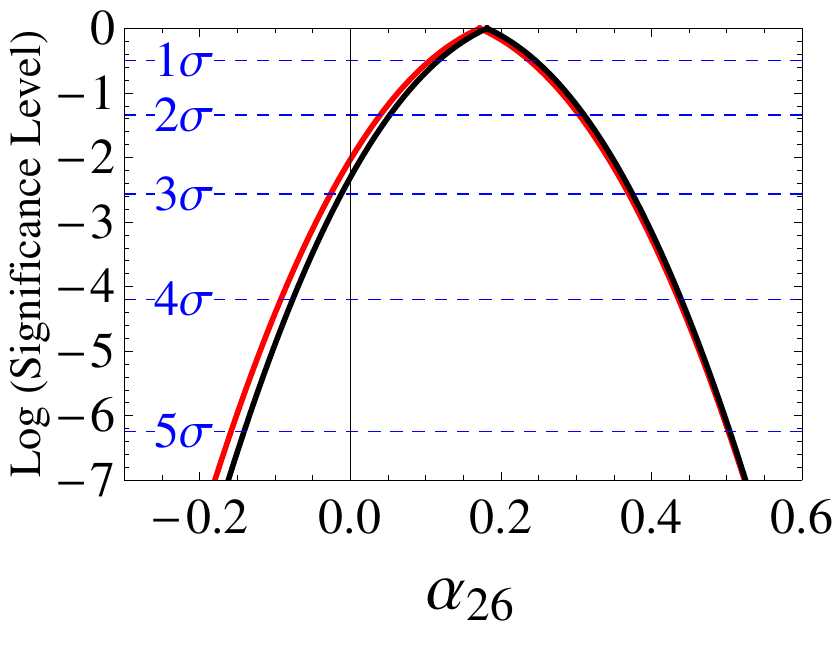}
	\hspace{1cm}
	\includegraphics[width=0.35\linewidth]{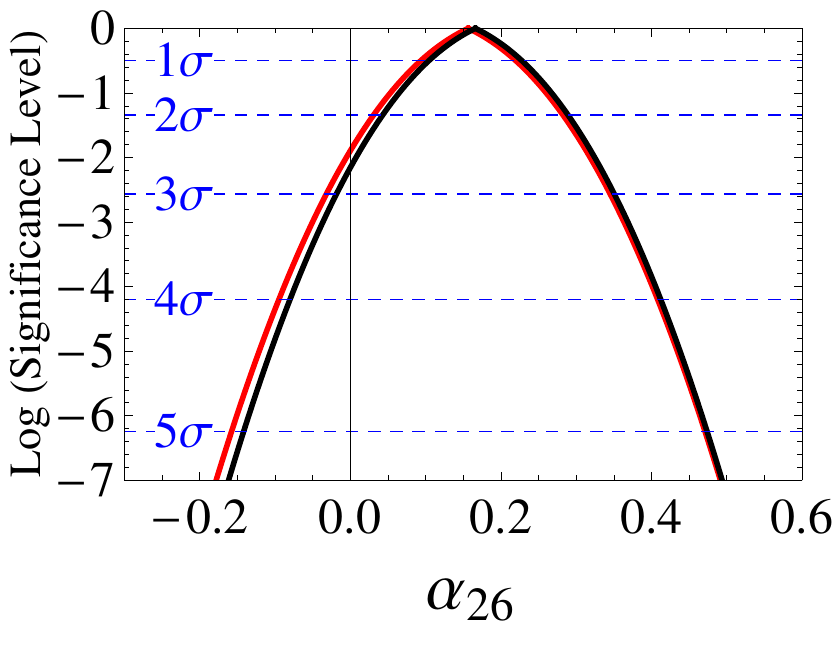}
	\caption{Global fits for $ \alpha_{26} $ from WD pulsation, WDLF, and RGB (cfr. table~\ref{tab:g13_global_fit}).
		The black (red) curves show the analysis which includes (does not include) the hint from the R-parameter (HB-hint). 
		\textit{Left panel:} assuming that the analyzed WD DA mode is trapped; \textit{Right panel:} removing the WD DA mode trapped assumption. 
	}
	\label{fig:gae2_global_fit}
\end{figure}

To quantitatively compare the different hints to $ \alpha_{26} $ shown in Tab.~\ref{tab:g13_squared} we preformed a $ \chi^2 $ analysis, where each hint is weighted with its own $ 1\,\sigma $ error.
In our analysis, however, we did not consider the two results from the DAV as independent hints. 
Although the two measurements were independently performed, the two stars share very similar properties and both results were based on a very similar hypothesis. 
Perhaps, the most conservative procedure should be to consider R548 as the representative for the DAV, since it has a larger error. 
In this case we find the results shown in Tab.~\ref{tab:g13_global_fit}.
The significance level of the combined analysis is shown in Fig.~\ref{fig:gae2_global_fit}. 
\begin{table}[h]
	\begin{center}
		\begin{tabular}{ c   c  l  p{5cm} |}\hline\hline
			mode trapped		& ~~HB hint~~	&  results 	\\ \hline	
			YES					&		NO				&  $ \chi^2_{\rm min}=2.96~ $(3 d.o.f.), 	\\	
			 					&						& $ g_{ae}=0$ excluded at $ 2.6\,\sigma \,, $ \\ 
			 					&						& $ \alpha_{26}=0.17\pm 0.13 ~ ({\rm at~ 2}\,\sigma) \,, $ \\ 
			 					&						& $ g_{ae}\left(\times 10^{13}\right)=1.47^{+0.48}_{-0.76}~ ({\rm at~ 2}\,\sigma) \,, $ \\ 
			 					&						&  \\
			NO 					&		NO				& $ \chi^2_{\rm min}=0.57~ $(3 d.o.f.),  	\\	 
			 					&						& $ g_{ae}=0$ excluded at $ 2.5\,\sigma \,, $ \\ 
			 					&						& $ \alpha_{26}=0.16\pm 0.13 ~ ({\rm at~ 2}\,\sigma) \,, $ \\ 
								&						& $ g_{ae}\left(\times 10^{13}\right)=1.41^{+0.48}_{-0.78}~ ({\rm at~ 2}\,\sigma)\,, $  \\ 
			 					&						&  \\
			YES					&		YES				&  $ \chi^2_{\rm min}=3.42~ $(4 d.o.f.), 	\\	
			 					&						& $ g_{ae}=0$ excluded at $ 2.8\,\sigma \,, $ \\ 
			 					&						& $ \alpha_{26}=0.18\pm 0.13 ~ ({\rm at~ 2}\,\sigma) \,, $ \\ 
			 					&						& $ g_{ae}\left(\times 10^{13}\right)=1.50^{+0.47}_{-0.71}~ ({\rm at~ 2}\,\sigma) \,, $ \\ 
			 					&						&  \\
			NO 					&		YES				& $ \chi^2_{\rm min}=1.11~ $(4 d.o.f.),  	\\	 
			 					&						& $ g_{ae}=0$ excluded at $ 2.7\,\sigma \,, $ \\ 
			 					&						& $ \alpha_{26}=0.17\pm 0.12 ~ ({\rm at~ 2}\,\sigma) \,, $ \\ 
								&						& $ g_{ae}\left(\times 10^{13}\right)=1.46^{+0.45}_{-0.67}~ ({\rm at~ 2}\,\sigma)\,. $  \\ \hline
		\end{tabular}
		\caption{Results for $g_{ae} $ at $ 1\,\sigma $ from WD pulsation, WDLF and RG stars.}
		\label{tab:g13_global_fit}   
	\end{center}
\end{table}

Notice that using the result from G117 - B15A, rather than R548, would give a somewhat worse fit ($ \chi^2_{\rm min}=10.5 $) perhaps indicating that the errors provided in this case are too optimistic or that the mode is indeed not trapped, or simply that ALPs are not the correct solution for this problem.

\M{Finally, note that all the considerations above apply to massless HPs with a magnetic coupling to electrons~\eqref{magneticHP}. The amplitudes of emission of an ALP and a massless HP from non-relativistic electrons are proportional after angular averaging~\cite{Hoffmann:1987et,Raffelt:1996wa}. Defining 
\begin{equation}
\alpha_{V26} = 2\frac{1}{4\pi}\(\frac{4 m_e v}{\Lambda^2} \)^2 10^{26}\,,
\end{equation}
where the factor of 2 accounts for the two HP polarisations, the emission rate of HPs is identical to that of an ALP coupled to electrons, with the substitution $\alpha_{26}\to \alpha_{V26}$. In particular, the HP emission can fit the WD, RG and HB anomalies as efficiently as an ALP coupled to electrons. Note also that the best fits $\alpha_{V26}\sim 0.17$ correspond to a new energy scale $\Lambda \sim 2$ PeV. However, the new physics can  appear at lower energies if the operator~\eqref{magneticHP} has an additional chirality suppression, i.e. $1/\Lambda^2 \sim \lambda_e/M^2$ where $\lambda_e$ is the electron Yukawa coupling. In this case $M\sim 3$ TeV and new physics could be at reach at the LHC~\cite{Dobrescu:2004wz}.  }

\section{Conclusions}
\label{sec:Conclusion}

For several years various independent observations of different stellar systems have shown an anomalous efficient cooling.
The current situation is summarized in Fig.~\ref{fig:WisePlot}.

\begin{figure}[htbp]
\begin{center}
\includegraphics[width=10cm]{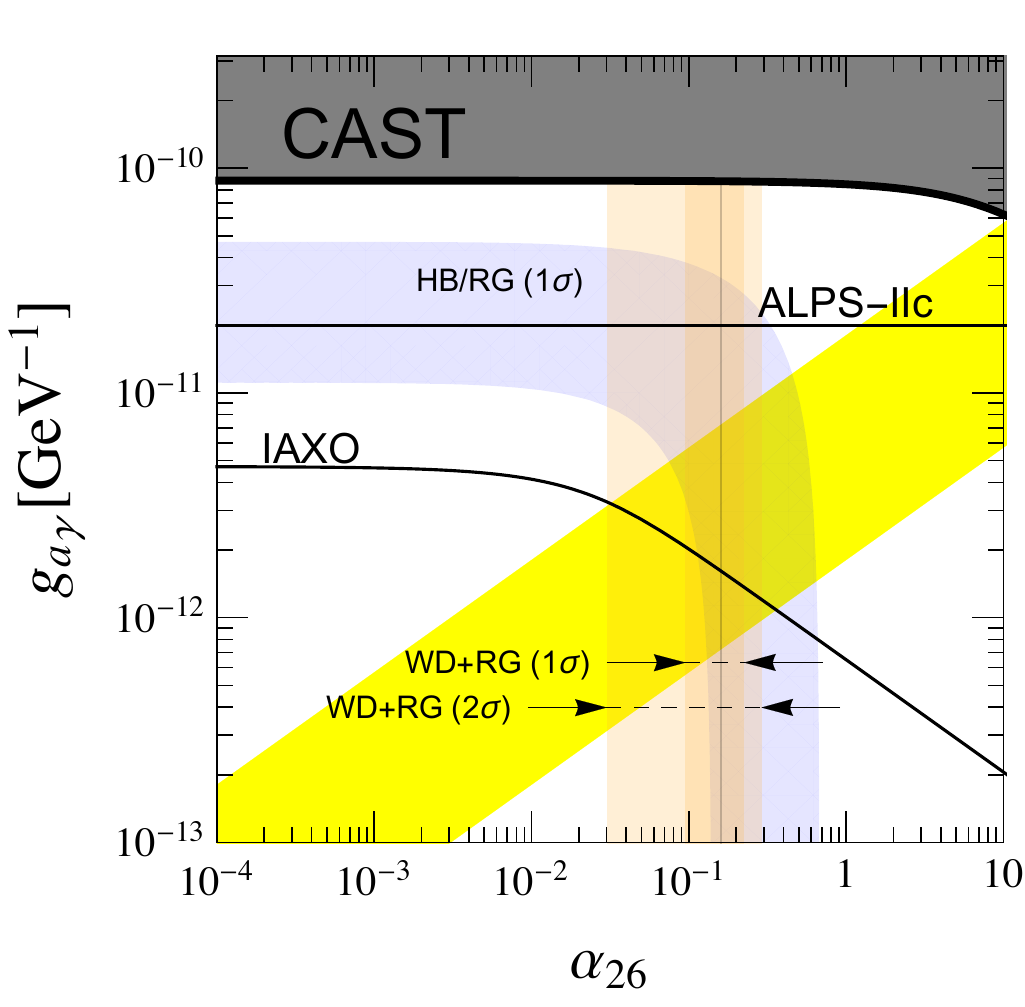}
\caption{Summary of the hinted region in the ALP parameter space (ALP-photon/ALP-electron coupling).
See text for details. 
The yellow band indicates motivated QCD axion models.
The projected sensitivity of the next generation ALP experiments, ALPS II and IAXO, is also shown. }
\label{fig:ALP_summary}
\end{center}
\end{figure}

In this work we have presented a coherent analysis of these anomalies and of the possible \textit{new physics} solutions
based on WISPs. 
Our analysis shows a particular preference for a pseudoscalar axion-like solution.
A summary of the hinted regions in the ALP-photon/ALP-electron parameter space is shown in Fig.~\ref{fig:ALP_summary}, where the light blue band labeled HB/RG is derived assuming the more recent results in~\cite{OscarPatras} for the $ R $-parameter (cf. sec.~\ref{sec:Hints to the ALP-electron coupling from WD and RG}).

The other WISPy candidates studied, on the other hand, seem to be inadequate to explain the combined observed cooling excesses. 
As discussed in sec.~\ref{sec:The case of a neutrino magnetic moment}, a neutrino magnetic moment would be a good candidate to explain the observation of the RGB tip, but fails to improve the WDLF or to explain the longer than expected rate of period change in the DAV. 
Additionally, it would be inadequate in explaining the hinted faster than expected cooling of the HB stars. 

The \M{massive~} hidden photon solution, though appealing since HPs could explain some of the observed anomalies and have a peculiar emission rate which could improve the WDLF fit, seem to require mass and coupling in regions already excluded by other astrophysics arguments (see Fig.~\ref{fig:summary_HP}). 
Additionally, there  does not seem to be a good overlap of the hinted areas.

The ALP solution, on the other hand, seems to be particularly efficient.

First of all, observations of the cold section of the WDLF, which can be approximated as a power law and so allow for a model independent analysis of new physics processes, show a preference for an additional cooling channel characterized by an emission rate of the form $ T^n $, with $ n\sim 4 $, the exponent predicted for the ALP production rate in a WD.

Additionally, the combined analysis of the WDs and RGB stars hints to an axion-electron coupling $ g_{ae}\simeq1.5\times 10^{-13} $, while the non-ALP solution seems to be excluded at more than 2.5$\, \sigma $, as shown in Tab.~\ref{tab:g13_global_fit}.
As shown in Fig.~\ref{fig:ALP_summary}, large portions of the hinted area is within reach of the proposed ALP search experiments, ALPS II~\cite{Bahre:2013ywa} and IAXO~\cite{Irastorza:2011gs,Armengaud:2014gea,Vogel:2015yka}.

Last but not least, if appropriately coupled to photons and nucleons, ALPs could also explain the observed anomalous $ R $-parameter~\cite{Ayala:2014pea} and the fast cooling of the NS in Cassiopeia A~\cite{Leinson:2014ioa}.
In fact, it has been noted in Ref. \cite{Ringwald:2015lqa}, that all the cooling excesses and even the transparency hint can be explained by a  
pseudo Nambu-Goldstone boson ALP with  
$m_a\lesssim 0.1\ {\rm \mu eV}$
and symmetry breaking scale 
\begin{equation}
f_a\sim 10^8\ {\rm GeV},
\end{equation} 
provided that the dimensionless coefficients relating the respective coupling strength and $f_a$, 
\begin{equation}
g_{a\gamma}=\frac{\alpha}{2\pi}\frac{C_{a\gamma}}{f_a}, \hspace{3ex}
g_{ae}=\frac{C_{ae}m_e}{f_a},\hspace{3ex}
g_{an}=\frac{C_{an}m_n}{f_a},
\end{equation}
are of order\footnote{ 
Intriguingly, in LARGE volume string compactifications, in which the volume $\mathcal V$ of the compactified extra dimensions is stabilized at an exponentially large value in units of the string size,
the decay constant of closed string ALPs is generically much smaller than the Planck scale,  $f_a \sim M_{\rm Pl}/\sqrt{\mathcal V}$, and their matter coupling coefficients $C_{af}$   are generically 
suppressed by a factor $\alpha\sim 10^{-2}$ in comparison to the photon coupling coefficient $C_{a\gamma}$, 
realizing the required properties \cite{Cicoli:2012sz}.}
\begin{equation} 
C_{a\gamma}\sim 1,\hspace{6ex} C_{ae}\sim C_{an}\sim 10^{-2}.
\end{equation}
Alternatively, the stellar cooling excesses  may be explained by a QCD axion $A$ with 
\begin{equation}
f_A \sim 10^9\ {\rm GeV}, \hspace{3ex} m_A\sim {\rm meV}, \hspace{3ex}C_{A\gamma}\sim C_{Ae}\sim C_{An}\sim 1,
\end{equation}
while the transparency hint requires then an additional ALP with 
\begin{equation}
f_a \sim 10^8\ {\rm GeV}, \hspace{3ex} m_a\lesssim 0.1\,{\rm \mu eV},\hspace{3ex} C_{a\gamma}\sim 1,
\hspace{3ex} C_{ae}\sim C_{an}\ll 1.
\end{equation}
Intriguingly, a QCD axion of DFSZ type in the above mass range could even account for the dark matter in 
the universe \cite{Ringwald:2015dsf}.   
A more detailed analysis of specific well motivated models which could account for such parameters is in preparation.

\M{Note finally that a massless hidden photon would provide another appealing solution. 
In fact, though such a field would have a quite different UV completion, its low energy phenomenology associated with electron interactions due to dimension 6 operators  \cite{Dobrescu:2004wz} would be, up to a constant scaling factor, identical to that of an axion~\cite{Hoffmann:1987et}. 
Therefore, a massless HP could address the WD, RGB and R-parameter hints as efficiently  as an ALP coupled to electrons. 
The new energy scale in this case can be a few TeV and testable in particle colliders.
Note however that the R-parameter  fit prefers also some additional cooling for HB stars which could not be provided by a massless HP. 
 }

Though a novel cooling channel induced by a new particle is not the only solution to the cooling anomalies, it may be the simplest, if we are willing to accept new physics. 
Additionally, it is certainly an appealing solution. 
After the negative results of the LHC searches for a massive dark matter candidate, perhaps we should look elsewhere, to the low energy frontier, and stars may be hinting at its presence.

\section*{Acknowledgements}

We acknowledge interesting discussions with Miller Bertolami, Dieter Horns, Axel Lindner, Alexandre Payez and Maxim Pospelov. 
\M{We also wish to thank the anonymous referee for constructive critical comments that improved the discussion. }
J. R. acknowledges support from the Ram\'on y Cajal fellowship RYC-2012-10957.


\end{document}